\documentclass[%
 reprint,
 superscriptaddress,
 preprintnumbers,
 amsmath, amssymb,
 aps,
 prd,
 nofootinbib,
 floatfix,
]{revtex4-2}

\usepackage{fancyhdr}

\usepackage{aas_macros}
\usepackage{bm}
\usepackage{graphicx}
\usepackage{array, makecell}

\usepackage{xcolor}
\definecolor{plasmablue}{rgb}{0.050383, 0.029803, 0.527975}
\usepackage{hyperref}
\hypersetup{
    colorlinks=true,
    linkcolor=plasmablue,
    urlcolor=plasmablue,
    citecolor=plasmablue,
    pdftitle={Multiprobe Cosmology from the Abundance of SPT Clusters and DES Galaxy Clustering and Weak Lensing},
    pdfauthor={Bocquet, Grandis, Krause, To et al. (the DES and SPT collaborations)},
    }

\renewcommand{\vec}[1]{\mbox{\boldmath$#1$}}

\newcommand{\LCDM}{\ensuremath{\Lambda\mathrm{CDM}}}
\newcommand{\wCDM}{\ensuremath{w\mathrm{CDM}}}
\newcommand{\threextwopt}{3$\times$2pt}
\newcommand{\Om}{\ensuremath{\Omega_\mathrm{m}}}
\newcommand{\sumMnu}{\ensuremath{\sum m_\nu}}
\newcommand{\sig}{\ensuremath{\sigma_8}}

\begin{document}

\pagestyle{fancy}
\fancyhf{}
\fancyhead[L]{\textsc{Bocquet et al.}}
\fancyhead[C]{\textsc{Multiprobe Cosmology from DES and SPT}}
\fancyhead[R]{\thepage}

\preprint{DES-2024-861}
\preprint{FERMILAB-PUB-24-779-PPD}

\title{Multiprobe Cosmology from the Abundance of SPT Clusters\\and DES Galaxy Clustering and Weak Lensing}

\author{S.~Bocquet\textsuperscript{1}}
\author{S.~Grandis\textsuperscript{2}}
\author{E.~Krause\textsuperscript{3}}
\author{C.~To\textsuperscript{4,5,6}}
\author{L.~E.~Bleem\textsuperscript{7,6}}
\author{M.~Klein\textsuperscript{1}}
\author{J.~J.~Mohr\textsuperscript{1,8}}
\author{T.~Schrabback\textsuperscript{2,9}}
\author{A.~Alarcon\textsuperscript{7,10}}
\author{O.~Alves\textsuperscript{11}}
% ...

\collaboration{the DES and SPT Collaborations}
\noaffiliation

\makeatletter
\def\Dated@name{}
\makeatother
\date{Phys. Rev. D accepted Jan 2025. Affiliations at the end of the paper. Contact author: \url{sebastian.bocquet@physik.lmu.de}}

\begin{abstract}
Cosmic shear, galaxy clustering, and the abundance of massive halos each probe the large-scale structure of the Universe in complementary ways. We present cosmological constraints from the joint analysis of the three probes, building on the latest analyses of the lensing-informed abundance of clusters identified by the South Pole Telescope (SPT) and of the auto- and cross-correlation of galaxy position and weak lensing measurements (3$\times$2pt) in the Dark Energy Survey (DES). We consider the cosmological correlation between the different tracers and we account for the systematic uncertainties that are shared between the large-scale lensing correlation functions and the small-scale lensing-based cluster mass calibration. Marginalized over the remaining $\Lambda$ cold dark matter ($\Lambda$CDM) parameters (including the sum of neutrino masses) and 52 astrophysical modeling parameters, we measure $\Omega_\mathrm{m}=0.300\pm0.017$ and $\sigma_8=0.797\pm0.026$. Compared to constraints from \textit{Planck} primary cosmic microwave background (CMB) anisotropies, our constraints are only 15\% wider with a probability to exceed of 0.22 ($1.2\sigma$) for the two-parameter difference. We further obtain $S_8\equiv\sigma_8(\Omega_\mathrm{m}/0.3)^{0.5}=0.796\pm0.013$ which is lower than the \textit{Planck} measurement at the $1.6\sigma$ level. The combined SPT cluster, DES 3$\times$2pt, and \textit{Planck} datasets mildly prefer a nonzero positive neutrino mass, with a 95\% upper limit $\sum m_\nu<0.25$~eV on the sum of neutrino masses. Assuming a $w$CDM model, we constrain the dark energy equation of state parameter $w=-1.15^{+0.23}_{-0.17}$ and when combining with \textit{Planck} primary CMB anisotropies, we recover $w=-1.20^{+0.15}_{-0.09}$, a $1.7\sigma$ difference with a cosmological constant. The precision of our results highlights the benefits of multiwavelength multiprobe cosmology and our analysis paves the way for upcoming joint analyses of next-generation datasets.
\end{abstract}

\maketitle

\section{Introduction}
The standard model of cosmology, the Lambda cold dark matter model (\LCDM), describes the cosmic expansion history and the growth of cosmic structure and is consistent with a variety of datasets.
One key pillar in testing this cosmological model is the accurate tracing of structure growth from the early Universe at redshift $z\sim1100$, when the cosmic microwave background (CMB) radiation was released, to the late-time Universe $z\lesssim2$.
For over a decade, such studies have been limited by the relatively weak constraining power of late-time Universe datasets, whereas primary CMB anisotropy measurements, in particular as measured by \textit{Planck}, have exhibited tighter constraints.
With the advent of wide-field lensing and galaxy surveys such as the Dark Energy Survey (DES) \citep{flaugher15, DES16, DES18DR1}, the Kilo-Degree Survey (KiDS) \citep{dejong12}, and the Hyper Suprime-Cam Subaru Strategic Program (HSC SSP) \citep{aihara18}, however, the constraining power of local probes is boosted and is about to reach the regime where the clustering properties of matter at $z\lesssim1$ can be determined as precisely as those at $z\sim1100$.

A key probe of the matter power spectrum enabled by these surveys is the measurement of the three possible two-point correlation functions (hence \threextwopt) among the galaxy position and galaxy weak lensing fields.
Such measurements have provided tight cosmological constraints \citep{heymans21, DESY33x2pt, more23hsc, miyatake23hsc, sugiyama23hsc} with a tantalizing hint that the late-time Universe may not have the exact properties as expected from interpreting the \textit{Planck} data in the context of the \LCDM\ model:
The parameter $S_8$ tends to be somewhat low in the late-time Universe constraints.\footnote{$S_8\equiv\sigma_8\sqrt{\Om/0.3}$ is a combination of the amplitude of fluctuations in the linear matter density field $\sigma_8$ and the matter density \Om. In the two-dimensional \Om--\sig\ plane, $S_8$ is the combination that is best constrained by current cosmic shear analyses.}

Another key probe of late-time structure formation on megaparsec scales is the abundance of massive halos and of the galaxy clusters they host.
While clusters can be identified in optical data, a particularly robust and well-understood selection scheme consists in observing the halos' hot intracluster medium, which emits X-rays via Bremsstrahlung and which causes a spectral distortion of the CMB via the thermal Sunyaev-Zel'dovich effect (hereafter SZ) \citep{sunyaev&zeldovich72}.
High-resolution, deep millimeter-wave surveys of the CMB enabled the first blind detection of a galaxy cluster in 2009 \citep{staniszewski09} and by now have enabled the detection of thousands of massive clusters out to $z\lesssim2$, e.g., \cite{bleem15, planck16XXVII, hilton21, bleem24, klein24ACT, zubeldia24SZiFi}.
To turn the abundance of SZ-selected clusters into a cosmological probe, one needs to relate the strength of the SZ signature to the underlying halo mass (we refer to this exercise as ``mass calibration'').
The halo mass creates the link to the halo mass function and thus the cosmological parameters and model.
A particularly robust means of mass calibration is based on measurements of weak gravitational shear around clusters, which can be modeled with exquisite control over systematic uncertainties.
Indeed, the analysis of SZ-selected clusters discovered in data from the South Pole Telescope (SPT) \citep{carlstrom11} with weak-lensing mass calibration using DES and \textit{Hubble} Space Telescope (HST) data enables competitive cosmological constraints \citep{bocquet24II}. In the \Om--\sig\ parameter space, the difference of these constraints relative to the DES Year~3 (Y3) \threextwopt\ results has a probability to exceed (PTE) of 0.25 ($1.1\sigma$).\footnote{The PTE is the probability of obtaining a larger difference between two measurements or between a measurement and a model prediction than what is observed. A very low PTE would thus imply that the observed difference is larger than random chance would allow given the uncertainties.}

In this paper, we present a joint analysis of the abundance of SPT-selected galaxy clusters and galaxy clustering and weak-lensing two-point correlation functions measured in the DES~Y3 dataset.
As we will demonstrate explicitly, the two probes are essentially independent even if they are both based on the same lensing data and probe the same matter field.
Indeed, the SPT cluster mass calibration relies on weak-lensing shear profiles measured in DES~Y3 data, but because these profiles are restricted to small scales $r<3.2/(1+z_\mathrm{cluster})~h^{-1}$Mpc, and because the SPT analysis is still limited by statistical uncertainties, the correlation with the \threextwopt\ measurements on larger scales is negligible.
Therefore, our joint analysis enables significantly tighter constraints than obtained from the probes individually.

The first joint analysis of the abundance of \emph{optically} selected galaxy clusters, cluster mass calibration from \emph{large-scale} cluster lensing, cluster clustering, and cluster-galaxy clustering, and \threextwopt\ \citep{y1-6x2+N} used data from the 1,321~deg$^2$ DES Year~1 survey.
A sample of 4,794 galaxy clusters \citep{redmapper} was analyzed jointly with weak-lensing \citep{y1-shapes,y1-bpz} and galaxy clustering \citep{y1-redmagic} data, building on the DES Year~1 \threextwopt\ analysis \citep{y1-3x2pt}.
The analysis presented here complements the DES~Year~1 multiprobe analysis as we use a different cluster selection, mass calibration, and inference methodology. 

In Sec.~\ref{sec:data}, we review the \threextwopt\ and cluster datasets and summarize the respective analysis frameworks.
In Sec.~\ref{sec:methods}, we discuss the correlations between the two datasets and how we join the two analyses.
We present the results in Sec.~\ref{sec:results} and conclude with a summary in Sec.~\ref{sec:summary}.

\section{Data and individual analysis frameworks}
\label{sec:data}
We use data from the first three years of DES (covering nearly 5,000~deg$^2$) and from the first two SPT surveys (SPT-SZ and SPTpol, covering a total of nearly 5,200~deg$^2$), along with additional cluster follow-up data as described below.
The two surveys share a common patch of 3,567~deg$^2$.

\subsection{DES Y3 \threextwopt}

We use the DES Y3 \threextwopt\ dataset and analysis as described in \cite{DESY33x2pt} and references therein.
Briefly, the \threextwopt\ analysis combines weak-lensing measurement of 100 million ``source'' galaxies \citep{y3-shapes} and positions of 10.7 million ``lens'' galaxies with magnitude-limited selection (\textsc{MagLim}) \citep{y3-maglim}.
The lensing shear measurements are obtained in a data-driven way by estimating the response to artificial shear applied to the images using the \textsc{metacalibration} algorithm \citep{metacal}.
To enable tomography, the sources are split in four redshift bins and the lenses are split in six redshift bins, of which only four are used in the analysis.

The data vector $\bm d$ contains the three two-point correlation function measurements and shear ratio measurements on small scales between source redshift bins that share the same lens bin \citep{sanchez:prat:22shearratio}.
The corresponding theory vector $\bm t_\mathrm{M}$ is computed for a given model M.
In this model, the nonlinear matter power spectrum is computed using \textsc{halofit} \citep{smith:peacock:jenkins:03, takahashi12halofit}.
Intrinsic alignment of (source) galaxies is modeled with the tidal alignment and tidal torquing (TATT) model \citep{blazek19alignment}, which is an extension of the nonlinear linear alignment (NLA) model.\footnote{The NLA model describes intrinsic alignment as a linear function of the nonlinear matter power spectrum, hence the name.}
The theory vector $\bm t_\mathrm{M}$ depends on the parameters $\vec p$.
These include the cosmological parameters, but also 29 nuisance parameters that describe, e.g., the lens galaxy bias, the shear and photo-$z$ calibrations, intrinsic alignment, etc.\footnote{Note that of the 29 nuisance parameters of the DES~Y3 \threextwopt\ model, only those describing the calibration of the effective source redshift distribution and residual shear biases may be correlated with the SPT cluster lensing analysis (see Sec.~\ref{sec:sharedsys}).}
The likelihood $\mathcal L$ is assumed to be Gaussian and we write
\begin{equation}
  \label{eq:threextwopt_likelihood}
  \begin{split}
    \ln \mathcal L(\bm d|\vec p, \mathrm{M}) =& -\frac12 \Bigl[\bm d -\bm t_\mathrm{M}(\vec p)\Bigr]^\mathrm{T} \,\bm{\mathrm{C}}^{-1}\, \Bigl[\bm d -\bm t_\mathrm{M}(\vec p)\Bigr] \\
    &+ \mathrm{const.}
  \end{split}
\end{equation}
with the covariance matrix $\bm{\mathrm{C}}$ that is computed analytically \citep{friedrich21covmat}.
We exactly follow the DES Y3 \threextwopt\ analysis and modeling choices as published and hence defer to \cite{DESY33x2pt} for details.

\subsection{SPT (SZ+pol) cluster cosmology}

The SPT cluster cosmology dataset and analysis framework are described in \cite{bocquet24I}.
Cluster candidates are identified by applying a matched filter to data from the 2,500~deg$^2$ SPT-SZ, the 2,800~deg$^2$ SPTpol Extended Cluster Survey (ECS), and the 500~deg$^2$  SPTpol 500d surveys and measuring the detection significance $\xi$ \citep{bleem15, bleem20, bleem24}.
Over the footprint of the SPT survey that is shared with DES, we consider detections with $\xi>4.25$ for SPTpol 500d, $\xi>4.5$ for SPT-SZ, and $\xi>5$ for SPTpol ECS.
We perform the cluster confirmation and redshift assignment using the multicomponent matched filter algorithm (MCMF) \cite{klein18, klein24spt}.
We use DES data, and at high redshift $z>1.1$, data from the Wide-field Infrared Survey Explorer (WISE) \cite{WISEobservatory}.
In essence, a candidate is confirmed as a cluster if the measured richness $\lambda$ exceeds a redshift-dependent limit $\lambda_\mathrm{min}(z)$ that is empirically calibrated to ensure a target purity of $>98\%$.
Outside of the shared survey footprint, we consider detections with $\xi>5$ and perform the cluster confirmation and redshift determination based on targeted follow-up observations (using among others, the PISCO imager \cite{stalder14PISCO} and \textit{Spitzer}/IRAC \citep{fazio04}) as described in \cite{bleem15, bleem20}.

The cosmology sample comprises 1,005 confirmed clusters at $z>0.25$.
A subset of 688 clusters at $z<0.95$ also have DES~Y3 weak-lensing data \citep{bocquet24I}, and 39 clusters at higher redshifts of 0.6--1.7 have weak-lensing data from HST \citep{schrabback18, raihan20, hernandez-martin20, schrabback21, zohren22}.
The lensing measurements exclude the cluster core regions and are restricted to the well-understood small-scale 1-halo term regime.
For DES, we consider scales between $0.5~h^{-1}$Mpc and $3.2/(1+z)~h^{-1}$Mpc, and for HST, scales between 0.5~Mpc and 1.5~Mpc.
The individual radial profiles of the tangential shear $\vec g_\mathrm{t}$, along with a ``lensing mass to halo mass'' relation $M_\mathrm{WL}-M_\mathrm{halo}$ that accounts for all stochastic and systematic uncertainties \citep{grandis21}, are used to calibrate the mean observable--mass relation.
We describe the cluster sample and the weak-lensing data using a Bayesian hierarchical model M as
\begin{equation}
  \label{eq:cluster_likelihood}
  \begin{split}
    \ln \mathcal L&\left(\{\xi_i,\lambda_i,z_i,\vec g_{\mathrm{t},i}\}_{i=1}^{N_\mathrm{cluster}}\big|\vec p, \mathrm{M}\right) =\\
    & \sum_{i=1}^{N_\mathrm{cluster}} \ln\frac{\mathrm{d}^4 N_\mathrm{M}(\vec p)}{\mathop{\mathrm{d}\xi} \mathop{\mathrm{d}\lambda} \mathop{\mathrm{d} \vec g_\mathrm{t}} \mathop{\mathrm{d} z}}
    \Big|_{\xi_i, \lambda_i, \vec g_{\mathrm{t},i}, z_i} \\
    &- \idotsint \mathop{\mathrm{d}\xi} \mathop{\mathrm{d}\lambda} \mathop{\mathrm{d} \vec g_\mathrm{t}} \mathop{\mathrm{d} z} 
     \frac{\mathrm{d}^4 N_\mathrm{M}(\vec p)}{\mathop{\mathrm{d}\xi} \mathop{\mathrm{d}\lambda} \mathop{\mathrm{d} \vec g_\mathrm{t}} \mathop{\mathrm{d} z}} \Theta_\mathrm{s}(\xi,\lambda,z) \\
    &+\mathrm{const.}
  \end{split}
\end{equation}
with the sample selection $\Theta_\mathrm{s}(\xi,\lambda,z)$ and where the index $i$ runs over all clusters in the sample. 
The differential cluster abundance is computed as
\begin{equation}
  \label{eq:dN}
  \begin{split}
    \frac{\mathrm{d}^4 N_\mathrm{M}(\vec p)}{\mathop{\mathrm{d}\xi} \mathop{\mathrm{d}\lambda} \mathop{\mathrm{d} \vec g_\mathrm{t}} \mathop{\mathrm{d} z} } =&
  \int \mathop{\mathrm{d}\Omega_\mathrm{s}} \idotsint \mathop{\mathrm{d} M} \mathop{\mathrm{d}\zeta} \mathop{\mathrm{d}\tilde\lambda} \mathop{\mathrm{d} M_\mathrm{WL}} \\
  &P(\xi|\zeta)
  P(\lambda|\tilde\lambda)
  P(\vec g_\mathrm{t}|M_\mathrm{WL}, \vec p) \\
  &P(\zeta, \tilde\lambda, M_\mathrm{WL} |M,z,\vec p) \\
  &\frac{\mathrm{d}^2 N(M, z, \vec p)}{\mathop{\mathrm{d} M} \mathop{\mathrm{d} V}} \frac{\mathrm{d}^2 V(z,\vec p)}{\mathop{\mathrm{d} z} \mathop{\mathrm{d}\Omega_\mathrm{s}}}
  \end{split}
\end{equation}
with the halo mass function \citep{tinker08} $\frac{\mathrm{d}^2 N(M, z, \boldsymbol{p})}{\mathop{\mathrm{d} M} \mathop{\mathrm{d} V}}$ and the differential volume $\frac{\mathrm{d}^2 V(z,\boldsymbol p)}{\mathop{\mathrm{d} z} \mathop{\mathrm{d}\Omega_\mathrm{s}}}$ within the survey footprint $\Omega_\mathrm{s}$.
The second line in Eq.~(\ref{eq:dN}) contains the relationships between the observed (and thus noisy) cluster properties and the intrinsic ones.
Finally, $P(\zeta, \tilde\lambda, M_\mathrm{WL} |M,z,\vec p)$ describes the multiobservable-to-mass scaling relations, including the effects of correlated intrinsic scatter.
For the cluster analysis, the vector $\vec p$ contains the cosmological parameters along with 23 parameters that describe the observable--mass relations and importantly, the systematic uncertainties in the weak-lensing modeling.

\section{Analysis Method}
\label{sec:methods}

In this section, we describe how we join the DES~Y3 \threextwopt\ and SPT cluster abundance analyses.
We note that the two analyses were performed blindly to avoid confirmation bias.
Because we strictly follow the analysis choices and modeling frameworks of the existing analyses, we do not need to (and cannot) perform a blind analysis.

We summarize our joint analysis as follows, and refer the reader to the individual subsections for further details.
We compute the cross-covariance between the two datasets due to the coupling of long-range modes in the matter density field.
Given the current size of the cluster sample and the current magnitude of the shot and shape noise in the cluster lensing measurements, we find that the cross-covariance does not contribute significantly and is ignored in the analysis that follows.
Therefore, we can simply sum the existing log-likelihood functions [Eqs.~(\ref{eq:threextwopt_likelihood}) and (\ref{eq:cluster_likelihood})], but we account for the fact that some of the lensing systematics (in particular, the uncertainty on the source redshift distribution) are shared between the \threextwopt\ analysis and the cluster mass calibration.
We do so by imposing a correlation between the respective parameters [see Eq.~(\ref{eq:wcorr}) below].
Instead of sampling the high-dimensional parameter space (six \LCDM\ parameters, 29 nuisance parameters for \threextwopt, 23 parameters for the cluster observable--mass relations), we importance sample the respective posterior parameter distributions.
To overcome the inherent noise in importance sampling, we train normalizing flows from which we can draw sufficiently large numbers of samples to obtain our final constraints.
Finally, we assess the quality of the joint fit and conclude that the mean recovered model is an adequate description of the data.

\subsection{Impact of cross-covariance between the SPT cluster abundance and DES \threextwopt}
\label{sec:cov}

Galaxy clusters trace the peaks of the large-scale structure. Hence, the abundance of galaxy clusters and halo-scale cluster mass profiles (probed by small-scale cluster lensing) are inherently correlated with tracers of the cosmic density field. In turn, the cross-covariance of cluster and \threextwopt\ measurements in the same survey footprint and with overlapping redshift ranges and scales is nonzero.
Note that in our analysis, however, the expected level of cross-covariance is small because the angular scales tested by the two probes are different. In this section, we demonstrate that we can safely ignore this cross-covariance given the level of uncertainties in the current measurements.
Neglecting the cross-covariance has the practical advantage that no additional development for a joint analysis pipeline is needed and that we can instead keep utilizing the existing ones.

First, we verify that for the SPT cluster dataset, the uncertainties in cluster lensing are dominated by shape noise and the uncertainties in the abundance are dominated by shot noise.
Adopting a halo model approach \citep{Cosmolike2017, 2021MNRAS.502.4093T}, we analytically calculate the covariance matrix of the cluster abundance and the stacked cluster lensing data. This calculation assumes that clusters are separated into two bins in SPT detection significance $\xi$ and three redshift bins.
The cluster sample selection also involves a cut in optical richness, the so-called ``optical cleaning,'' e.g., \citep{klein18, klein24spt}.
For the SPT cluster sample, the impact of optical cleaning is small, and for simplicity, we ignore it here (but we do account for it in the cluster likelihood).
We generate the simulated data vector $\bm d_\mathrm{sim}$ and the covariance matrix $\bm\Sigma$ using the scale cuts for cluster lensing as in the SPT analysis \citep{bocquet24I} and assuming the best-fit $\xi$--mass relation and cosmological values obtained from that analysis \citep{bocquet24II}.
We then calculate the signal-to-noise ratio 
\begin{equation}
    \mathrm{SNR}\equiv \sqrt{\bm d_\mathrm{sim}^\mathrm{T}\bm\Sigma^{-1}\bm d_\mathrm{sim}}
\end{equation}
of the simulated data with the full covariance matrix and with the covariance matrix that only contains shape and shot noise terms. We find that the signal-to-noise ratios of the two covariance matrices differ at the $\sim 3\%$ level, which would have minimal impact on the cosmological constraints.
We note that the SPT cluster lensing analysis \citep{bocquet24I}, and thus also this work, does not perform a stacked analysis; instead, it considers each cluster individually in a hierarchical Bayesian likelihood framework. This is equivalent to an analysis with infinitely small $\xi$ and redshift bins. To apply our stacked result to \citep{bocquet24II}, we verify the sensitivity of our calculation to the number of $\xi$ and redshift bins. Specifically, we perform another set of stacked analyses by increasing the number of bins by a factor of 6 and do not see a difference in our result. We thus conclude that our result also applies to the analysis framework adopted in \citep{bocquet24II}.

The fact that the cluster lensing and abundance data vectors are dominated by shot and shape noise already justifies ignoring the cross-covariance between the cluster data vector and the \threextwopt\ data vector in our combined analysis.
However, we explicitly test the impact of ignoring this cross-covariance term on our combined analyses. Still using the halo model approach, we calculate the full covariance matrix of \threextwopt{}, cluster abundance, and cluster lensing. We then calculate the signal-to-noise ratio of the whole \threextwopt{} and cluster data vector using the full covariance matrix and using the covariance matrix without cross terms of \threextwopt{} and cluster parts. We find that the differences in signal-to-noise ratios are at the $\sim 0.05\%$ level, solidifying our conclusion that we can safely assume that the cluster dataset and \threextwopt{} are independent of each other.

\subsection{Determination of shared systematics}
\label{sec:sharedsys}

As the DES weak-lensing measurements of SPT clusters and the shear two-point correlation functions that enter the DES \threextwopt\ data vector use the same lensing source galaxy shapes and photo-$z$s, systematic uncertainties in these properties impact both cosmological probes in a correlated way.
Note that there is no correlation between the DES galaxy position two-point correlation function and the SPT cluster abundance.
Similarly, the 39 cluster lensing measurements based on HST data are not correlated with the DES lensing dataset.
For the DES~Y3 \threextwopt\ analysis, the lensing source galaxies were split in four tomographic bins according to their mean redshift estimates.
For each bin $b$, a mean redshift bias $\Delta z_\mathrm{s}^b$ and a mean multiplicative shear bias $m^b$ were determined, along with the systematic uncertainties on both quantities (eight parameters in total) \citep{myles21des, maccrann22des}.
The SPT cluster lensing analysis used the same source selection, although tomographic bin 1 was dropped entirely \citep{bocquet24I}.
The systematic uncertainty in the cluster weak-lensing mass calibration was determined by calibrating a ``weak-lensing mass to halo mass'' relation [$M_\mathrm{WL}-M_\mathrm{halo}$, see Eqs.~(36)--(38) in \cite{bocquet24I}] using Monte Carlo simulations of synthetic cluster lensing measurements based on mass maps from hydrodynamical simulations (following \cite{grandis21}).
In these Monte Carlo simulations, the lensing source photo-$z$s and the shear bias parameters $m$ were stochastically drawn from the calibrated distributions, thereby incorporating the effects of the uncertain photo-$z$ and shear calibration into the uncertainties in the $M_\mathrm{WL}-M_\mathrm{halo}$ relation.

In this work, we repeat the calibration of the cluster lensing model, but for each Monte Carlo realization, we now also record the shear bias $m^b$ and the uncertainty on the mean redshift $\Delta z_\text{s}^b$ for each tomographic bin $b$.
This allows us to track the correlation between the parameters of the $M_\mathrm{WL}-M_\mathrm{halo}$ model and $\Delta z_\text{s}^b$ and $m^b$.
We determine that only the first principal component of the cluster weak-lensing mass bias $b_\mathrm{WL}$ and the mean redshift bias of the fourth tomographic redshift bin $\Delta z_\mathrm{s}^4$ (anti)correlate significantly, with a correlation coefficient $\rho=-0.81$.\footnote{Note that the first principal component of the cluster lensing mass bias is defined as $\sigma_{\ln b_{\mathrm{WL},1}}$ in the SPT analysis \citep{bocquet24I,bocquet24II} but we use a shorter notation $b_\mathrm{WL}$ here.}
The negative correlation is explained as follows: If, for example, the source redshift is biased low, then, for a given lensing signal, the inferred lensing mass would be biased high, and with it, the amplitude of the $M_\mathrm{WL}-M_\mathrm{halo}$ relation.
All other parameters exhibit negligible levels of correlation.
This is expected because only at relatively large cluster redshifts does the uncertainty in the photo-$z$ calibration represent a significant contribution to the overall systematic error budget (see discussion and Fig.~10 in \cite{bocquet24I}). The uncertainty in the shear calibration is negligible at all cluster redshifts.
In our analysis, we account for the correlation $\rho$ as discussed in the next subsection.

We note that the characterization of the shared systematics between \threextwopt\ and the weak-lensing cluster mass calibration as performed here is straightforward because of the deliberate choice taken in the SPT cluster cosmology analysis to perform the cluster weak-lensing analysis based on the same source selection as used for the DES lensing two-point correlation functions.
While alternative, more optimal cluster lensing source selection schemes would almost certainly have led to slightly reduced statistical uncertainties, the characterization of the systematic uncertainties and, in particular, the characterization of their correlation with the systematic uncertainties in \threextwopt\ would have been more complicated.
Therefore, we recommend a similar analysis philosophy also for future multiprobe analyses that include weak-lensing calibrated cluster abundance measurements.

\subsection{Parameter inference}
\label{sec:IS}

We follow the DES~Y3 \threextwopt\ analysis and apply the same uniform priors on \Om, $\Omega_\mathrm{b}$, $\Omega_\nu h^2$, $h$, $n_s$, and $A_s$, see Table~I in \cite{DESY33x2pt}.
We consider \sig\ as a derived parameter.

As demonstrated in the previous subsections, the cluster abundance and mass calibration likelihood and the \threextwopt\ likelihood are effectively independent.
Therefore, instead of running an excessively expensive Markov
chain Monte Carlo (MCMC) to explore the joint high-dimensional parameter space, we adopt an importance sampling approach.
In this approach, the samples of the posterior parameter distribution of one analysis are updated by multiplying their weights with the likelihood of the other analysis; the resulting samples describe the joint distribution.
Typically though, this procedure leads to noisy posterior distributions because the effective sample size decreases.
To mitigate this effect, we first train normalizing flows to learn the posterior distributions of the SPT cluster and DES \threextwopt\ analyses.\footnote{The original SPT cluster analysis assumed different priors on the cosmological parameters \citep{bocquet24II}. To enable the importance sampling analysis, we reran the cluster analysis using the DES~Y3 priors.
The recovered results were essentially unchanged.}

A normalizing flow is a generative model in machine learning that learns the bijective mapping between a simple distribution and the target probability distribution \citep{rezende15NF, kobyzev20NFintro, papamakarios17MAF}.
In our case, the simple distribution is a multivariate normal distribution, and the training set are the MCMC samples (from the SPT or DES analysis).
Once the transformation is known, one can draw a large number of samples from the target distribution by drawing samples from the normal distribution and applying the transformation.
Furthermore, one can obtain the posterior probability at any point in parameter space by applying the inverse transformation.
We use a modified implementation of \textsc{FlowJax}\footnote{\url{https://danielward27.github.io/flowjax}.} that can handle weighted samples.\footnote{Our implementation is available at \url{https://github.com/SebastianBocquet/flowjax}.}

In this work, we are interested in the improvements on the cosmological parameter constraints enabled by the joint analysis.
Therefore, we restrict the importance sampling to the parameters \Om, $\Omega_\mathrm{b}$, $\Omega_\nu h^2$, $h$, $n_s$, \sig, and the correlated nuisance parameters $b_\mathrm{WL}$ and $\Delta z_\mathrm{s}^4$.
This reduces the dimensionality of the parameter space to eight and improves the stability of the normalizing flows and of the importance sampling analysis.
In Appendix~\ref{sec:app_IS}, we demonstrate that the trained flows are able to accurately reproduce the true distributions (see the upper-right triangle in Fig.~\ref{fig:IS}).
We now draw a large number of samples from one flow and update the sample weights $w$ using the likelihood at that location in parameter space from the other flow.
Finally, we account for the correlation between $b_\mathrm{WL}$ and $\Delta z_\mathrm{s}^4$ (see previous section) by updating the sample weights
\begin{equation}
  \label{eq:wcorr}
  \begin{split}
    \bm \delta \equiv & \begin{pmatrix}b_\mathrm{WL}-\langle b_\mathrm{WL}\rangle\\\Delta z_\mathrm{s}^4-\langle \Delta z_\mathrm{s}^4\rangle\end{pmatrix},\\
    \bm \Sigma_\mathrm{uncorr.} \equiv& \begin{pmatrix}\sigma^2_{b_\mathrm{WL}}&0\\0&\sigma^2_{\Delta z_\mathrm{s}^4}\end{pmatrix}, \\
    \bm \Sigma_\mathrm{corr.} \equiv&
      \begin{pmatrix}
        \sigma^2_{b_\mathrm{WL}} & \rho\,\sigma_{b_\mathrm{WL}}\,\sigma_{\Delta z_\mathrm{s}^4} \\
        \rho\,\sigma_{b_\mathrm{WL}}\,\sigma_{\Delta z_\mathrm{s}^4} & \sigma^2_{\Delta z_\mathrm{s}^4}
      \end{pmatrix}, \\
    \ln w_\text{with corr.} =& \ln w_\text{no corr.} 
    + \frac12\bm \delta^\mathrm{T}
    \bm \Sigma_\mathrm{uncorr.}^{-1}
    \bm \delta
    - \frac12\bm \delta^\mathrm{T}
    \bm \Sigma_\mathrm{corr.}^{-1}
    \bm \delta.
  \end{split}
\end{equation}
We use the importance sampled chains to extract the parameter constraints presented in this work.
Our baseline results are based on the DES~Y3 \threextwopt\ chain, updated with the SPT cluster likelihood and after applying Eq.~(\ref{eq:wcorr}).
These constraints are shown in solid red in Fig.~\ref{fig:IS}.
To cross-check that the importance sampling scheme is robust, we also extract results starting for the SPT cluster chain, importance sampling using the DES~Y3 \threextwopt\ likelihood, and applying Eq.~(\ref{eq:wcorr}), as shown in red dashed lines in Fig.~\ref{fig:IS}.
The two analysis routes lead to almost indistinguishable results, confirming the reliability of our approach.

\begin{figure*}
  \includegraphics[width=.49\textwidth]{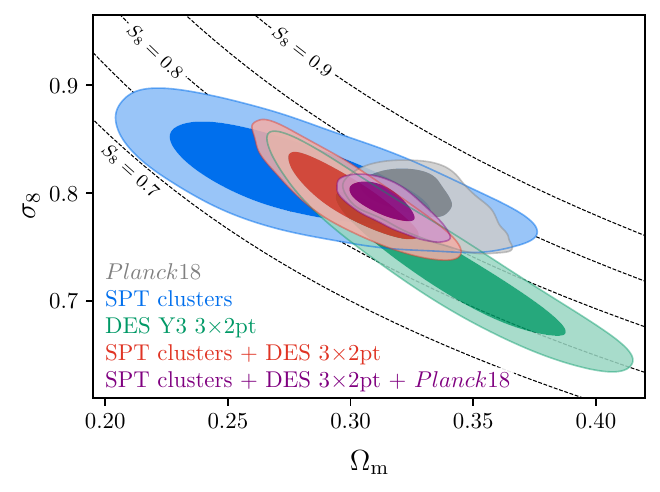}
  \hfill
  \includegraphics[width=.49\textwidth]{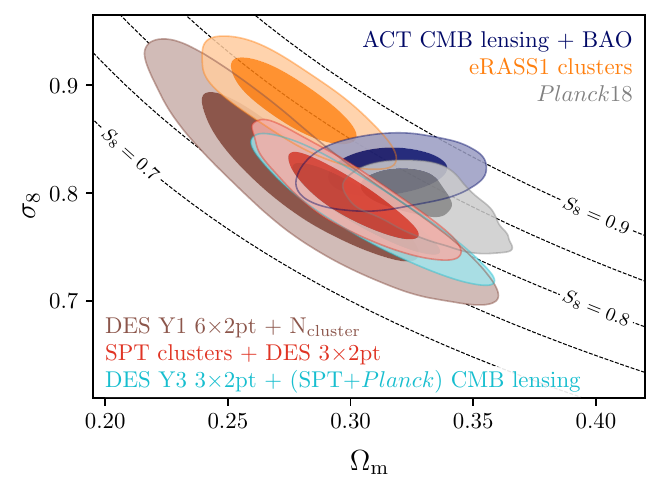}
  \caption{Constraints on \Om\ and $\sigma_8$ (68\% and 95\% credibility) in \LCDM\ with massive neutrinos.
  The two panels show the same parameter ranges.
  Dashed lines show lines of constant $S_8\equiv\sigma_8\sqrt{\Om/0.3}$.
  Left: the individual lensing-informed SPT cluster abundance and DES \threextwopt\ results, along with their combination.
  We also show the constraints from \textit{Planck} TT,TE,EE+lowE primary CMB anisotropies and the combination with our joint analysis.
  Right: comparison of our SPT clusters + DES \threextwopt\ results with a selection of external single-probe and multiprobe analyses.}
  \label{fig:Omegam-sigma8}
\end{figure*}

\subsection{Goodness of fit}

We test whether the best-fit model for the joint cluster and \threextwopt\ analysis is an adequate description of the data.
The \threextwopt\ data vector contains 471 data points.
The $\chi^2$ at the maximum \emph{a posteriori} probability (MAP) of the joint analysis ($\chi^2=538.5$) is higher than the $\chi^2$ at the MAP of the \threextwopt\ \LCDM\ analysis by $\Delta\chi^2=1.8$.
The implied PTE decreases by about 10\%.
Note that if the MAP of the joint analysis is not identical to the MAP of the individual probe (and if the priors are not changed) then a somewhat worse fit is to be expected by definition.
The \threextwopt\ analysis does not show signs of significant internal inconsistencies (PTE of 0.023 \citep{DESY33x2pt, doux21}).
Because the PTE in the joint analysis is only slightly lower, we conclude that the \threextwopt\ data are well fit in our joint analysis, too.
We also compare the total number of clusters and the measured stacked shear profiles with the model predictions at the MAP of the joint analysis (analogous to Figs.~1 and 2 in \cite{bocquet24II}) and obtain $\chi^2=43.1$ for 27 data points, with a corresponding PTE of 0.03 ($\chi^2=35.6$ for the clusters-only analysis).
Note that the cluster analysis is performed using an unbinned Poisson likelihood for the cluster sample and a hierarchical Bayesian likelihood for the individual cluster lensing shear profiles. Stacked data are only used to evaluate the goodness of fit.
We conclude that the model is an adequate description of the joint dataset and present the recovered cosmological constraints.

\section{Results}
\label{sec:results}

We present our constraints on \LCDM, the sum of neutrino masses, and \wCDM.
Throughout this work, we assume spatial flatness and a uniform prior $[0.06,\,0.6]$~eV for the sum of neutrino masses (the lower limit is given by measurements of neutrino oscillations, see e.g., \cite{PDG2024}).

\subsection{\LCDM}

Figure~\ref{fig:Omegam-sigma8} (left) shows \LCDM\ constraints in the \Om--\sig\ plane as obtained by SPT clusters, DES~\threextwopt, and our joint analysis SPT clusters + DES~\threextwopt.
The constraints on a selection of parameters are presented in Table~\ref{tab:results}.
The joint constraints lie at the intersection of the two individual probes.
The degeneracy direction mostly follows the degeneracy of the \threextwopt\ result; the parameter combination that is constrained with the smallest absolute uncertainty is $\sig(\Om/0.3)^{0.494}$, which is essentially $S_8$.
The ratio of the areas of the 95\% credible region in \Om--\sig\ space for SPT clusters, DES \threextwopt, and the joint analysis is $3.3:2.1:1$.
While the two probes cannot individually constrain the Hubble constant, the joint analysis breaks some of the parameter degeneracies and we recover $h=0.73\pm0.07$ (see also Fig.~\ref{fig:IS} in Appendix~\ref{sec:app_IS}).
However, this result is not strong enough to inform the Hubble tension.

In the \Om--\sig\ plane shown in Fig.~\ref{fig:Omegam-sigma8}, the 95\% credible region of the SPT cluster + DES \threextwopt\ analysis is 15\% larger than for \textit{Planck} 2018 TT,TE,EE+lowE \citep{planck18VI}.
We quantify the two-parameter difference with a PTE of $0.22$ ($1.2\sigma$).
Our measurement $S_8=0.796\pm0.013$ differs from the \textit{Planck} measurement $S_8=0.831\pm0.017$ at $1.6\sigma$.
We show lines of constant $S_8$ in the figure to help guide the eye.
Because the difference is not significant, we combine our joint constraints with \textit{Planck} temperature and polarization power spectra to obtain tight constraints on the cosmological parameters.
In Fig.~\ref{fig:Omegam-sigma8} (right), we also show the results from the multiprobe analyses of DES~Y1 cluster abundance, cluster clustering, galaxy clustering, and lensing \citep{y1-6x2+N} and of CMB lensing measured by the Atacama Cosmology Telescope (ACT) and baryon acoustic oscillations (BAOs) measured in the 6dF and SDSS galaxy surveys \citep{madhavacheril24ACT, qu24ACT}.
In the \Om--\sig\ plane, the ratio of the areas of the 95\% credible regions of the DES~Y1, ACT + BAO, and our SPT clusters + DES \threextwopt\ analyses is $2.8:1.0:1$.
Our results are similar to the joint analysis of DES~Y3 galaxy clustering and lensing and SPT+\textit{Planck} CMB lensing \citep{DESSPT22_6x2pt}, but somewhat tighter.

Comparing to the eROSITA eRASS1 cluster cosmology analysis \citep{ghirardini24}, if we assume that their analysis is independent from ours, we find that their reported value of \sig\ is higher than ours at the $2.4\sigma$ level, and we find a PTE of 0.018 ($2.4\sigma$) in the \Om--\sig\ plane.
In reality, however, the footprints of the DES, SPT, and eRASS1 surveys overlap and the eRASS1 and SPT cluster samples have objects in common, and both analyses rely on mass calibration using DES lensing data \citep{bocquet24I, grandis24erassDES}.
Carefully assessing the statistical significance of the difference is not the main goal of this study, and we leave this quantification for future works.

\begin{table*}
  \caption{Parameter constraints for the \LCDM\ and \wCDM\ models, marginalized over all cosmology and 52 nuisance parameters (mean and 68\% credible intervals, or 95\% limit).
  $\sig\left(\Om/0.3\right)^{0.5}$ is the parameter combination that is best constrained by \threextwopt, and $\sig\left(\Om/0.3\right)^{0.25}$ is the combination that is best constrained by the SPT cluster abundance.
  The cluster and \threextwopt\ datasets cannot individually constrain $h$ and we only quote the joint constraint (but we always marginalize over $h$).
  The joint analysis does not meaningfully constrain \sumMnu\ on its own, and we only quote the constraint obtained in combination with {\it Planck} 2018 TT,TE,EE+lowE.
  Note that while the \wCDM\ constraints from SPT clusters only are affected by the hard prior $w>-2$, the contours of the joint constraints close.
  \label{tab:results}}
  \begin{ruledtabular}
    \begin{tabular}{lccccccc}
      Dataset & \Om & \sig & $S_8\equiv\sig\left(\frac{\Om}{0.3}\right)^{0.5}$ & $\sig\left(\frac{\Om}{0.3}\right)^{0.25}$ & $h$ & \sumMnu~[eV] & $w$\\[1pt]
      \colrule\\[-8pt]
      \LCDM\\[1pt]
      SPT clusters & $0.286\pm0.032$ & $0.817\pm0.026$ & $0.795\pm0.029$ & $0.805\pm0.016$ & \dots & \dots & $-1$\\[2pt]
      DES \threextwopt & $0.339^{+0.032}_{-0.031}$ & $0.733^{+0.039}_{-0.049}$ & $0.776\pm0.017$ & $0.754\pm0.031$ & \dots & \dots & $-1$\\[3pt]
      \makecell[l]{SPT clusters\\~~+ DES \threextwopt} & $0.300\pm0.017$ & $0.797\pm0.026$ & $0.796\pm0.013$ & $0.796\pm0.017$ & $0.73\pm0.07$ & \dots & $-1$\\[6pt]
      \makecell[l]{SPT clusters\\~~+ DES \threextwopt\\~~+ \textit{Planck}} & $0.314\pm0.009$ & $0.791\pm0.013$ & $0.809\pm0.009$ & $0.800\pm0.010$ & $0.674\pm0.007$ & $0.14^{+0.02}_{-0.07} (<0.25)$ & $-1$\\
      \colrule\\[-8pt]
      \wCDM\\[1pt]
      SPT clusters & $0.268\pm0.037$ & $0.820\pm0.026$ & $0.772\pm0.040$ & $0.796\pm0.020$ & \dots & \dots & $-1.45\pm0.31$\\[2pt]
      DES \threextwopt & $0.352^{+0.035}_{-0.041}$ & $0.719^{+0.037}_{-0.044}$ & $0.775^{+0.026}_{-0.024}$ & $0.746\pm0.029$ & \dots & \dots & $-0.98^{+0.32}_{-0.20}$\\[3pt]
      \makecell[l]{SPT clusters\\~~+ DES \threextwopt} & $0.294\pm0.021$ & $0.793\pm0.023$ & $0.784\pm0.019$ & $0.788\pm0.015$ & $0.71\pm0.06$ & \dots & $-1.15^{+0.23}_{-0.17}$\\[6pt]
      \makecell[l]{SPT clusters\\~~+ DES \threextwopt\\~~+ \textit{Planck}} & $0.284\pm0.018$ & $0.811\pm0.020$ & $0.787\pm0.016$ & $0.799\pm0.013$ & $0.715\pm0.024$ & $0.25^{+0.07}_{-0.19}(<0.50)$ & $-1.20^{+0.15}_{-0.09}$
    \end{tabular}
  \end{ruledtabular}
\end{table*}

\subsection{Sum of neutrino masses}
\label{sec:summnu}

\begin{figure*}
  \includegraphics[width=\textwidth]{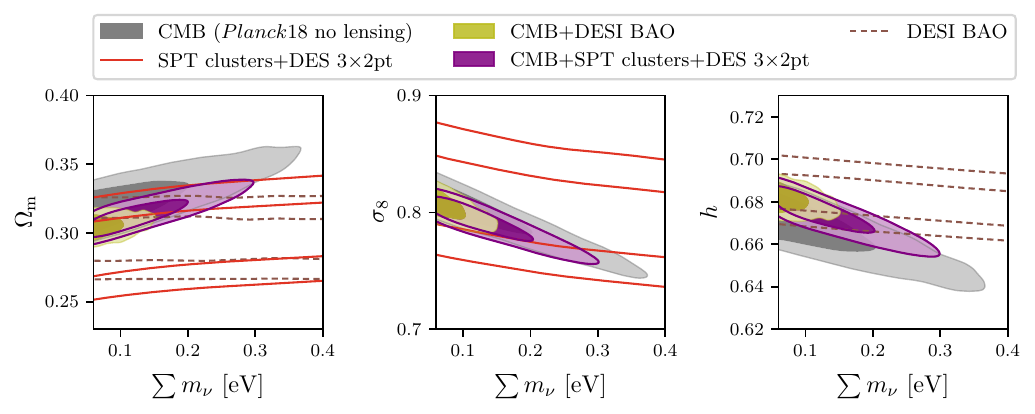}
  \caption{Joint constraints on the sum of neutrino masses \sumMnu\ and a selection of other cosmological parameters.
  For the sake of readability, we omit the comparatively weak constraints on $h$ from the SPT clusters + DES \threextwopt\ analysis, and we note that the BAO data do not constrain \sig.
  Neither the BAO data nor the joint SPT clusters + DES \threextwopt\ analysis can meaningfully constrain \sumMnu\ on their own.
  However, their combination with CMB data allows for refined constraints compared to those obtained from CMB data alone by breaking the \sumMnu--\Om\ degeneracy in the CMB-based constraints.
  The BAO data also break the \sumMnu--$h$ degeneracy, while the SPT clusters + DES \threextwopt\ analysis also breaks the \sumMnu--\sig\ degeneracy.
  }
  \label{fig:summnu2d}
\end{figure*}

\begin{figure}
  \includegraphics[width=\columnwidth]{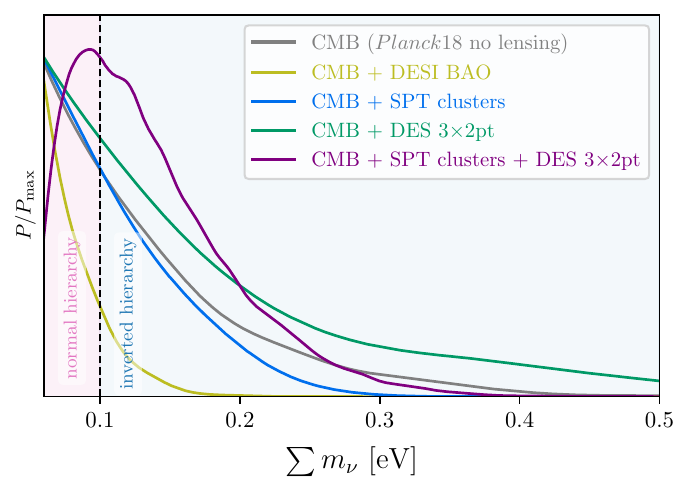}
  \caption{Marginalized constraints on the sum of neutrino masses. For the normal hierarchy, the minimal value is $\sumMnu>0.06$~eV, which is the prior we adopt in our analysis. For the inverted hierarchy, $\sumMnu>0.1$~eV, as indicated by the dashed line.
  The posterior of the joint \textit{Planck} + SPT clusters + DES \threextwopt\ analysis peaks at 0.09~eV, with no significant preference for either hierarchy.}
  \label{fig:summnu}
\end{figure}

Following the DES \threextwopt\ analysis, we set a lower limit $\sumMnu>0.06$~eV to reflect constraints from neutrino oscillations.
Measurements of \textit{Planck}18 TT,TE,EE+lowE primary CMB anisotropies place an upper limit $\sumMnu<0.30$~eV at 95\% credibility \citep{planck18VI}.
As shown with the gray contours in Fig.~\ref{fig:summnu2d}, this constraint is limited by degeneracies with \Om, \sig, and $h$.
Therefore, the combination of CMB data with independent measurements of these other cosmological parameters breaks (some of) these degeneracies and enables tighter constraints.
In Fig.~\ref{fig:summnu2d}, we show that while our joint SPT clusters + DES \threextwopt\ analysis cannot meaningfully constrain \sumMnu, it breaks the degeneracies with \Om\ and \sig\ in the CMB analysis.
In Fig.~\ref{fig:summnu}, we show the marginalized posterior probability distribution for the sum of neutrino masses.
The constraints from \textit{Planck}18 and from the combination of that dataset with either \threextwopt\ or the SPT cluster abundance peak at the minimum allowed mass.
However, the upper limits in all three analyses cannot rule out the inverted hierarchy, which would imply $\sumMnu>0.1$~eV.
Interestingly, due to the breaking of degeneracies in the nontrivial high-dimensional parameter space, the combination of \textit{Planck}18 and our joint SPT clusters + DES \threextwopt\ analysis results in a constraint on the sum of neutrino masses that peaks at a nonzero value of 0.09~eV (mean value is 0.14~eV, see also Table~\ref{tab:results}).
However, the credible intervals are still wide enough that both the normal and the inverted mass hierarchy are compatible with our results.

For comparison, in Fig.~\ref{fig:summnu2d}, we also show the combination of CMB and BAO data from the Dark Energy Spectroscopic Instrument (DESI) \citep{DESI24VI}.
While BAOs do not constrain \sig, they provide measurements on \Om\ and $h$ and thus break the degeneracies in the CMB analysis (see also, e.g., \cite{loverde:weiner:24}) in a way that is complementary to our SPT clusters + DES \threextwopt\ analysis.

\subsection{\wCDM}

\begin{figure}
  \includegraphics[width=\columnwidth]{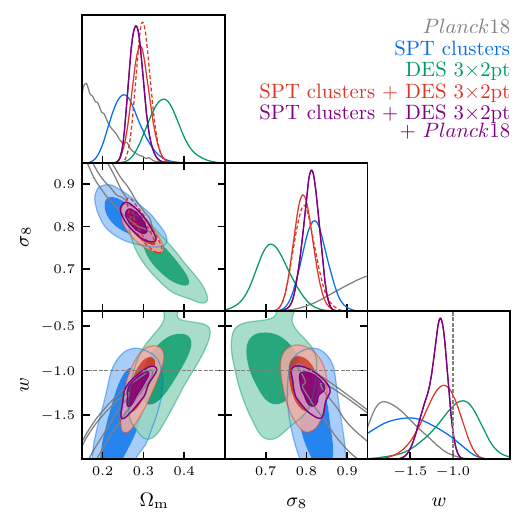}
  \caption{Constraints on the matter density, the amplitude of fluctuations, and the dark energy equation of state parameter (68\% and 95\% credibility) in \wCDM\ with massive neutrinos.
  For reference, we also show the \LCDM\ ($w=-1$) constraints on \Om\ and \sig\ from our SPT clusters + DES \threextwopt\ analysis with dashed lines.}
  \label{fig:wCDM}
\end{figure}

We now additionally allow the dark energy equation of state parameter $w$ to vary.
Figure~\ref{fig:wCDM} shows the constraints from DES \threextwopt, from the SPT cluster abundance, and from our joint analysis.
The constraints on \Om\ and \sig\ are comparable with the constraints recovered for the \LCDM\ model.
We report $w=-1.15^{+0.23}_{-0.17}$, which improves over the single-probe uncertainties by 35 and 22\%, and which agrees with a cosmological constant $w=-1$ with a PTE of 0.58 ($0.6\sigma$).
For reference, the purely geometrical measurement using DES Supernovae is yet another 26\% tighter and peaks at a less negative value $w=-0.80^{+0.14}_{-0.16}$. The DESI BAO measurement is also purely geometric and yields $w=-0.99^{+0.15}_{-0.13}$, which is 31\% tighter than our measurement and almost perfectly centered on $w=-1$ \citep{DESI24VI}.
Conversely, the results from \textit{Planck} data alone exhibit extended degeneracies between $w$ and many other parameters.
By combining \textit{Planck} temperature and polarization data with our SPT cluster + DES \threextwopt\ dataset, we can break these degeneracies and recover tight constraints; notably, our measurement $w=-1.20^{+0.15}_{-0.09}$ differs from a cosmological constant with a PTE of 0.09, or $1.7\sigma$ (see Table~\ref{tab:results} and purple contours in Fig.~\ref{fig:wCDM}).

The \textit{Planck} 2018 data are known to exhibit more smoothing at high multipoles than can be explained by lensing, and this can impact the recovered parameter constraints \citep{planck18VI}.
We investigate this effect by also allowing the amplitude of CMB lensing $A_\mathrm{L}$ to vary, and we recover $A_\mathrm{L}=1.20\pm0.06$ and $w=-1.16^{+0.16}_{-0.10}$.
The difference with a cosmological constant thus reduces to the $1\sigma$ level but $A_\mathrm{L}$ is greater than unity at more than $3\sigma$.

\section{Summary and Outlook}
\label{sec:summary}

In this work, we present a joint analysis of weak-lensing and galaxy clustering measurements from DES data and the abundance of SPT-selected clusters with DES and HST weak-lensing mass calibration.
The two individual probes have roughly comparable constraining power and we show that their cosmological correlation and the correlation due to using the same DES lensing dataset are both negligible.
Therefore, the joint analysis of these two probes is particularly appealing.

For a flat \LCDM\ model with massive neutrinos, we report competitive constraints on \Om\ and \sig\ (see Table~\ref{tab:results}).
In the two-parameter plane, the 95\% credibility region is only 15\% larger than the one allowed by \textit{Planck} 2018 primary CMB data (TT,TE,EE+lowE) \citep{planck18VI}.
We are thus witnessing the beginning of an era where measurements of the large-scale structure are (at least) as constraining as early-Universe CMB observations.
Our analysis does not provide a strong suggestion for $S_8$ being lower than measured by \textit{Planck}, but as with many other published results, our measurement lies below the \textit{Planck} value (at the $1.6\sigma$ level in our case).

The combined SPT cluster, DES \threextwopt, and \textit{Planck} dataset shows a mild preference for a nonzero positive sum of neutrino masses with an upper limit $\sumMnu<0.25$~eV.
Our joint analysis improves the constraints on the dark energy equation of state parameter $w$ over the results from the individual probes.
We recover $w=-1.15^{+0.23}_{-0.17}$ from SPT clusters + DES \threextwopt, and for the joint analysis with \textit{Planck}, $w=-1.20^{+0.15}_{-0.09}$.
However, these results cannot rival the existing constraints enabled by geometric probes such as BAO and Supernovae.

While dark energy with a time-evolving equation of state has seen renewed interest, we do not consider this model here.
Constraints on $w_0$ and $w_a$ were presented using the DES~Y3 \threextwopt\ data \citep{DESY3extensions} and their combination with DES Supernovae and SDSS BAO \citep{DESY5SNIa}.
However, the SPT Collaboration has not yet presented a $w_0w_a$CDM analysis using its cluster sample.
Therefore, we leave a joint SPT clusters and DES \threextwopt\ analysis of the $w_0w_a$CDM model to future work that will show whether the contours in $w_0-w_a$ space close, which would enable an independent cross-check of the Supernovae + BAO (+~CMB) constraints \citep{DESI24VI, pantheonpluscosmo, union3cosmo, DESY5SNIa}.

This work presents the second joint analysis of the cluster abundance and \threextwopt\ measurements.
Compared to the first analysis \citep{y1-6x2+N}, which used optically selected clusters and large-scale cluster--shear correlation functions, we use the SZ-selected SPT cluster sample and cluster lensing measurements in the small-scale, 1-halo term regime.
Therefore, our analysis is complementary to the existing work.
Our work paves the way for future joint analyses of larger SZ-selected cluster samples (from, e.g., SPT-3G \cite{sobrin22}, ACT \citep{hilton21, klein24ACT}, the Simons Observatory \citep{ade19SO}, and CMB-S4 \citep{abazajian16cmbs4}), improved lensing and galaxy clustering datasets (e.g., DES Year 6, \textit{Euclid} \citep{laureijs11, Euclid24overview}, and LSST \citep{lsst09sciencebook} obtained with the Vera~C. Rubin Observatory), and updated CMB lensing measurements (such as presented in, e.g., \cite{ge:millea:24}).

In these future analyses, the statistical uncertainties in the cluster abundance and cluster lensing measurements will be reduced and we expect the correlation due to using the same lensing data as \threextwopt\ to no longer be negligible.
The upcoming cluster sample will extend down to lower halo masses, implying that the sample (co)variance and the cross-covariance with \threextwopt\ and other large-scale structure probes might no longer be negligible, either.
Our work thus sets the stage for future, more complex analyses of joint probes that will enable us to probe the large-scale structure of the Universe with unprecedented constraining power.

\begin{acknowledgments}

This research was supported by the Ludwig-Maximilians-Universit\"at M\"unchen, the MPG Faculty Fellowship program, and the Excellence Cluster ORIGINS, which is funded by the Deutsche Forschungsgemeinschaft (DFG, German Research Foundation) under Germany's Excellence Strategy - EXC-2094-390783311.
The Innsbruck authors acknowledge support from the Austrian Research Promotion Agency (FFG) and the Federal Ministry of the Republic of Austria for Climate Action, Environment, Mobility, Innovation and Technology (BMK) via grants 899537 and 900565, and 911971.
EK is supported in part by Department of Energy grant DE-SC0020247 and the David and Lucile Packard Foundation. CT is supported by the Eric and Wendy Schmidt AI in Science Postdoctoral Fellowship, a Schmidt Futures program.
Work at Argonne National Laboratory was supported by the U.S. Department of Energy, Office of High Energy Physics, under Contract No. DE-AC02-06CH11357.

The South Pole Telescope program is supported by the National Science Foundation (NSF) through awards OPP-1852617 and OPP-2332483. Partial support is also provided by the Kavli Institute of Cosmological Physics at the University of Chicago.

Funding for the DES Projects has been provided by the U.S. Department of Energy, the U.S. National Science Foundation, the Ministry of Science and Education of Spain, 
the Science and Technology Facilities Council of the United Kingdom, the Higher Education Funding Council for England, the National Center for Supercomputing 
Applications at the University of Illinois at Urbana-Champaign, the Kavli Institute of Cosmological Physics at the University of Chicago, 
the Center for Cosmology and Astro-Particle Physics at the Ohio State University,
the Mitchell Institute for Fundamental Physics and Astronomy at Texas A\&M University, Financiadora de Estudos e Projetos, 
Funda{\c c}{\~a}o Carlos Chagas Filho de Amparo {\`a} Pesquisa do Estado do Rio de Janeiro, Conselho Nacional de Desenvolvimento Cient{\'i}fico e Tecnol{\'o}gico and 
the Minist{\'e}rio da Ci{\^e}ncia, Tecnologia e Inova{\c c}{\~a}o, the Deutsche Forschungsgemeinschaft and the Collaborating Institutions in the Dark Energy Survey. 

The Collaborating Institutions are Argonne National Laboratory, the University of California at Santa Cruz, the University of Cambridge, Centro de Investigaciones Energ{\'e}ticas, 
Medioambientales y Tecnol{\'o}gicas-Madrid, the University of Chicago, University College London, the DES-Brazil Consortium, the University of Edinburgh, 
the Eidgen{\"o}ssische Technische Hochschule (ETH) Z{\"u}rich, 
Fermi National Accelerator Laboratory, the University of Illinois at Urbana-Champaign, the Institut de Ci{\`e}ncies de l'Espai (IEEC/CSIC), 
the Institut de F{\'i}sica d'Altes Energies, Lawrence Berkeley National Laboratory, the Ludwig-Maximilians-Universit{\"a}t M{\"u}nchen and the associated Excellence Cluster ORIGINS, 
the University of Michigan, NSF's NOIRLab, the University of Nottingham, The Ohio State University, the University of Pennsylvania, the University of Portsmouth, 
SLAC National Accelerator Laboratory, Stanford University, the University of Sussex, Texas A\&M University, and the OzDES Membership Consortium.

Based in part on observations at Cerro Tololo Inter-American Observatory at NSF's NOIRLab (NOIRLab Prop. ID 2012B-0001; PI: J. Frieman), which is managed by the Association of Universities for Research in Astronomy (AURA) under a cooperative agreement with the National Science Foundation.

The DES data management system is supported by the National Science Foundation under Grant Numbers AST-1138766 and AST-1536171.
The DES participants from Spanish institutions are partially supported by MICINN under grants ESP2017-89838, PGC2018-094773, PGC2018-102021, SEV-2016-0588, SEV-2016-0597, and MDM-2015-0509, some of which include ERDF funds from the European Union. IFAE is partially funded by the CERCA program of the Generalitat de Catalunya.
Research leading to these results has received funding from the European Research Council under the European Union's Seventh Framework Program (FP7/2007-2013) including ERC grant agreements 240672, 291329, and 306478.
We  acknowledge support from the Brazilian Instituto Nacional de Ci\^encia e Tecnologia (INCT) do e-Universo (CNPq grant 465376/2014-2).

This manuscript has been authored by Fermi Research Alliance, LLC under Contract No. DE-AC02-07CH11359 with the U.S. Department of Energy, Office of Science, Office of High Energy Physics.
This research has made use of the SAO/NASA Astrophysics Data System and of \textsc{adstex}.\footnote{\url{https://github.com/yymao/adstex}}

\end{acknowledgments}

\appendix

\section{Impact of shared systematics}

\begin{figure}
  \includegraphics[width=\columnwidth]{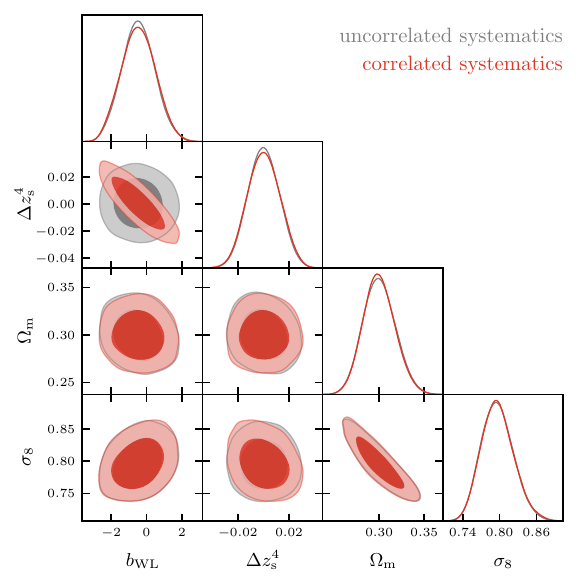}
  \caption{Impact of the shared systematics in the lensing source photo-$z$ calibration, which introduces a correlation $\rho=-0.81$ between the amplitude of the cluster lensing mass-to-halo mass $b_\mathrm{WL}$ and the uncertainty on the mean redshift of the lensing tomographic bin 4, $\Delta z_\mathrm{s}^4$.
  We show the 68\% and 95\% credibility regions.
  Neglecting the fact that the lensing systematics are shared between the two analyses (gray) has a negligible impact on the recovered cosmological constraints.
  Throughout this work, we nonetheless correctly account for the correlation (red).}
  \label{fig:corr}
\end{figure}

In our baseline analysis, we account for the correlation $\rho=-0.81$ between (one of) the parameters of the cluster lensing mass bias $b_\mathrm{WL}$ and the mean redshift bias of the fourth tomographic bin $\Delta z_\mathrm{s}^4$.
Here, we investigate the impact of this correlation.
In Fig.~\ref{fig:corr}, we show the two parameters along with the cosmological parameters of prime interest, \Om\ and \sig.
In the baseline analysis, the correlation between the two noncosmology parameters can be clearly seen.
For comparison, we also produce constraints without accounting for the correlation of lensing systematics, which we do by not applying Eq.~(\ref{eq:wcorr}).
As can be seen in the figure, ignoring the correlation between $b_\mathrm{WL}$ and $\Delta z_\mathrm{s}^4$ has a negligible impact on the recovered cosmological constraints.
This is expected, because the analyses of the individual probes are not limited by the uncertainty in the photo-$z$ calibration.
Nevertheless, in our baseline analysis, we properly take the correlation into account.

\section{Robustness of Normalizing Flows and Importance Sampling}
\label{sec:app_IS}

\begin{figure*}
  \includegraphics[width=\textwidth]{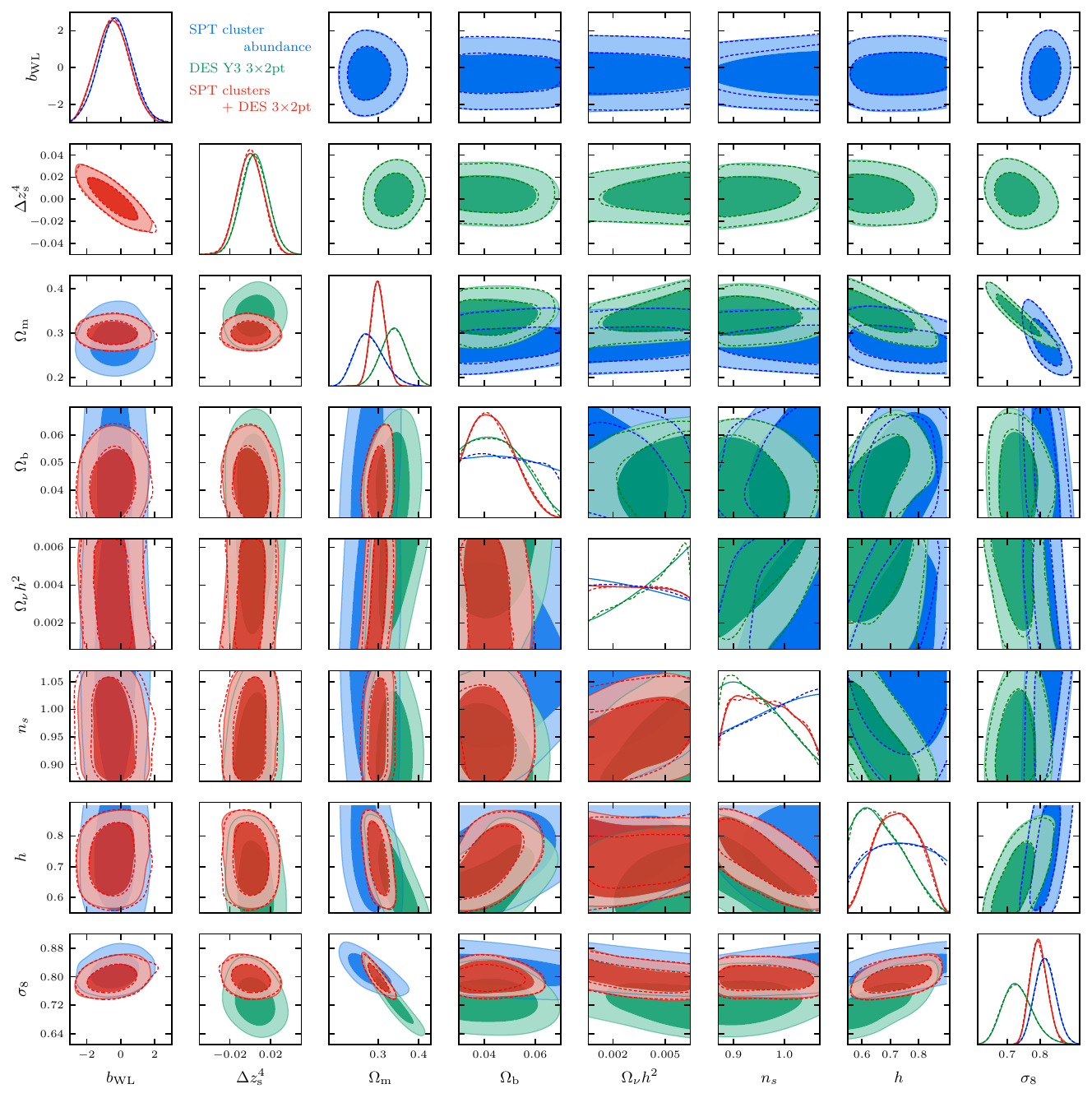}
  \caption{Parameter constraints (68\% and 95\% credibility) in \LCDM.
  Upper right triangle:
  The original SPT and DES analyses are shown in solid blue and green, posteriors obtained from the trained normalizing flows are shown with dashed lines.
  Lower left triangle:
  The fiducial joint analysis, in which the DES posterior is updated with the SPT likelihood, is shown with solid red lines; the inverse analysis (SPT posterior updated with DES likelihood) is shown in red dashed lines.
  }
  \label{fig:IS}
\end{figure*}

In Fig.~\ref{fig:IS}, we show the SPT cluster and DES \threextwopt\ chains, along with the posterior distributions obtained from the trained normalizing flows.
We observe that the well-constrained parameters such as the nuisance parameters and \Om\ and \sig\ are very well reproduced by the normalizing flows, whereas the reconstruction of the other parameters seems to be more challenging.
In the same figure, we also show our fiducial results, as obtained by importance sampling the \threextwopt\ results with the cluster likelihood (solid red lines and contours).
Finally, we also show the results obtained from the inverse approach, where we importance sample the probability distribution obtained from the cluster analysis with the \threextwopt\ likelihood (dashed red lines and contours).
There is very good qualitative agreement between the two sets of results, and we thus conclude that our inference scheme is robust.

\bibliography{paper}

\section*{Affiliations}
\noindent
\textsuperscript{1} University Observatory, Faculty of Physics, Ludwig-Maximilians-Universit\"at, Scheinerstr. 1, 81679 Munich, Germany\\
\textsuperscript{2} Universit\"at Innsbruck, Institut f\"ur Astro- und Teilchenphysik, Technikerstr. 25/8, 6020 Innsbruck, Austria\\
\textsuperscript{3} Department of Astronomy/Steward Observatory, University of Arizona, 933 North Cherry Avenue, Tucson, AZ 85721-0065, USA\\
\textsuperscript{4} Center for Cosmology and Astro-Particle Physics, The Ohio State University, Columbus, OH 43210, USA\\
\textsuperscript{5} Department of Astronomy and Astrophysics, University of Chicago, Chicago, IL 60637, USA\\
\textsuperscript{6} Kavli Institute for Cosmological Physics, University of Chicago, 5640 South Ellis Avenue, Chicago, IL 60637, USA\\
\textsuperscript{7} High-Energy Physics Division, Argonne National Laboratory, 9700 South Cass Avenue, Lemont, IL 60439, USA\\
\textsuperscript{8} Max Planck Institute for Extraterrestrial Physics, Gie\ss enbachstr.~1, 85748 Garching, Germany\\
\textsuperscript{9} Argelander-Institut f\"ur Astronomie, Auf dem H\"ugel 71, 53121 Bonn, Germany\\
\textsuperscript{10} Institute of Space Sciences (ICE, CSIC), Campus UAB, Carrer de Can Magrans, s/n, 08193 Barcelona, Spain\\
\textsuperscript{11} Department of Physics, University of Michigan, Ann Arbor, MI 48109, USA\\
\textsuperscript{12} Department of Astrophysical Sciences, Princeton University, Peyton Hall, Princeton, NJ 08544, USA\\
\textsuperscript{13} Institute for Astronomy, University of Hawai'i, 2680 Woodlawn Drive, Honolulu, HI 96822, USA\\
\textsuperscript{14} Physics Department, 2320 Chamberlin Hall, University of Wisconsin-Madison, 1150 University Avenue Madison, WI 53706-1390, USA\\
\textsuperscript{15} Department of Physics and Astronomy, University of Pennsylvania, Philadelphia, PA 19104, USA\\
\textsuperscript{16} Department of Physics, Northeastern University, Boston, MA 02115, USA\\
\textsuperscript{17} Brookhaven National Laboratory, Bldg 510, Upton, NY 11973, USA\\
\textsuperscript{18} Instituto de F\'isica Te\'orica, Universidade Estadual Paulista, S\~ao Paulo, Brazil\\
\textsuperscript{19} Laborat\'orio Interinstitucional de e-Astronomia - LIneA, Rua Gal. Jos\'e Cristino 77, Rio de Janeiro, RJ - 20921-400, Brazil\\
\textsuperscript{20} Department of Physics, Carnegie Mellon University, Pittsburgh, Pennsylvania 15312, USA\\
\textsuperscript{21} NSF AI Planning Institute for Physics of the Future, Carnegie Mellon University, Pittsburgh, PA 15213, USA\\
\textsuperscript{22} Instituto de Astrofisica de Canarias, E-38205 La Laguna, Tenerife, Spain\\
\textsuperscript{23} Universidad de La Laguna, Dpto. Astrofísica, E-38206 La Laguna, Tenerife, Spain\\
\textsuperscript{24} Center for Astrophysical Surveys, National Center for Supercomputing Applications, 1205 West Clark St., Urbana, IL 61801, USA\\
\textsuperscript{25} Department of Astronomy, University of Illinois Urbana-Champaign, 1002 W. Green Street, Urbana, IL 61801, USA\\
\textsuperscript{26} Physics Department, William Jewell College, Liberty, MO 64068, USA\\
\textsuperscript{27} Department of Physics, Duke University Durham, NC 27708, USA\\
\textsuperscript{28} NASA Goddard Space Flight Center, 8800 Greenbelt Rd, Greenbelt, MD 20771, USA\\
\textsuperscript{29} Jodrell Bank Center for Astrophysics, School of Physics and Astronomy, University of Manchester, Oxford Road, Manchester, M13 9PL, UK\\
\textsuperscript{30} Institut d'Estudis Espacials de Catalunya (IEEC), 08034 Barcelona, Spain\\
\textsuperscript{31} Kavli Institute for Particle Astrophysics and Cosmology, P. O. Box 2450, Stanford University, Stanford, CA 94305, USA\\
\textsuperscript{32} Lawrence Berkeley National Laboratory, 1 Cyclotron Road, Berkeley, CA 94720, USA\\
\textsuperscript{33} Fermi National Accelerator Laboratory, P.O. Box 500, Batavia, IL 60510, USA\\
\textsuperscript{34} Universit\'e Grenoble Alpes, CNRS, LPSC-IN2P3, 38000 Grenoble, France\\
\textsuperscript{35} Jet Propulsion Laboratory, California Institute of Technology, 4800 Oak Grove Dr., Pasadena, CA 91109, USA\\
\textsuperscript{36} Department of Physics \& Astronomy, University College London, Gower Street, London, WC1E 6BT, UK\\
\textsuperscript{37} Department of Physics and Astronomy, University of Waterloo, 200 University Ave W, Waterloo, ON N2L 3G1, Canada\\
\textsuperscript{38} California Institute of Technology, 1200 East California Boulevard, Pasadena, CA 91125, USA\\
\textsuperscript{39} Department of Astronomy, University of California, Berkeley, 501 Campbell Hall, Berkeley, CA 94720, USA\\
\textsuperscript{40} SLAC National Accelerator Laboratory, Menlo Park, CA 94025, USA\\
\textsuperscript{41} Kavli Institute for Cosmology, University of Cambridge, Madingley Road, Cambridge CB3 0HA, UK\\
\textsuperscript{42} Institut de F\'isica d'Altes Energies (IFAE), The Barcelona Institute of Science and Technology, Campus UAB, 08193 Bellaterra (Barcelona) Spain\\
\textsuperscript{43} School of Physics and Astronomy, Cardiff University, CF24 3AA, UK\\
\textsuperscript{44} Department of Astronomy, University of Geneva, ch. d'\'Ecogia 16, CH-1290 Versoix, Switzerland\\
\textsuperscript{45} Department of Physics, University of Arizona, Tucson, AZ 85721, USA\\
\textsuperscript{46} Department of Physics and Astronomy, Pevensey Building, University of Sussex, Brighton, BN1 9QH, UK\\
\textsuperscript{47} Instituto de Astrof\'isica e Ci\^encias do Espa\c co, Faculdade de Ci\^encias, Universidade de Lisboa, 1769-016 Lisboa, Portugal\\
\textsuperscript{48} Department of Applied Mathematics and Theoretical Physics, University of Cambridge, Cambridge CB3 0WA, UK\\
\textsuperscript{49} Perimeter Institute for Theoretical Physics, 31 Caroline St. North, Waterloo, ON N2L 2Y5, Canada\\
\textsuperscript{50} Instituto de F\'isica Gleb Wataghin, Universidade Estadual de Campinas, 13083-859, Campinas, SP, Brazil\\
\textsuperscript{51} Kavli Institute for the Physics and Mathematics of the Universe (WPI), UTIAS, The University of Tokyo, Kashiwa, Chiba 277-8583, Japan\\
\textsuperscript{52} Centro de Investigaciones Energ\'eticas, Medioambientales y Tecnol\'ogicas (CIEMAT), Madrid, Spain\\
\textsuperscript{53} Ruhr University Bochum, Faculty of Physics and Astronomy, Astronomical Institute, German Centre for Cosmological Lensing, 44780 Bochum, Germany\\
\textsuperscript{54} Nordita, KTH Royal Institute of Technology and Stockholm University, Hannes Alfv\'ens v\"ag 12, SE-10691 Stockholm, Sweden\\
\textsuperscript{55} Department of Physics, University of Genova and INFN, Via Dodecaneso 33, 16146, Genova, Italy\\
\textsuperscript{56} ICTP South American Institute for Fundamental Research \textbar\ Instituto de F\'isica Te\'orica, Universidade Estadual Paulista, S\~ao Paulo, Brazil\\
\textsuperscript{57} Space Telescope Science Institute, 3700 San Martin Drive, Baltimore, MD 21218, USA\\
\textsuperscript{58} Department of Physics and Astronomy, Stony Brook University, Stony Brook, NY 11794, USA\\
\textsuperscript{59} Institut de Recherche en Astrophysique et Plan\'etologie (IRAP), Universit\'e de Toulouse, CNRS, UPS, CNES, 14 Av. Edouard Belin, 31400 Toulouse, France\\
\textsuperscript{60} Excellence Cluster Origins, Boltzmannstr.\ 2, 85748 Garching, Germany\\
\textsuperscript{61} Department of Physics, Stanford University, 382 Via Pueblo Mall, Stanford, CA 94305, USA\\
\textsuperscript{62} Department of Physics, Southern Methodist University, Dallas, TX 75205, USA\\
\textsuperscript{63} Cerro Tololo Inter-American Observatory, NSF's National Optical-Infrared Astronomy Research Laboratory, Casilla 603, La Serena, Chile\\
\textsuperscript{64} Institute for Astronomy, University of Edinburgh, Edinburgh EH9 3HJ, UK\\
\textsuperscript{65} School of Physics and Astronomy, Cardiff University, Cardiff CF24 3YB, United Kingdom\\
\textsuperscript{66} Kavli Institute for Particle Astrophysics and Cosmology, Stanford University, 452 Lomita Mall, Stanford, CA 94305, USA\\
\textsuperscript{67} SLAC National Accelerator Laboratory, 2575 Sand Hill Road, Menlo Park, CA 94025, USA\\
\textsuperscript{68} School of Physics, University of Melbourne, Parkville, VIC 3010, Australia\\
\textsuperscript{69} NIST Quantum Devices Group, 325 Broadway Mailcode 817.03, Boulder, CO 80305, USA\\
\textsuperscript{70} Department of Physics, University of Colorado, Boulder, CO 80309, USA\\
\textsuperscript{71} Department of Physics, University of Cincinnati, Cincinnati, OH 45221, USA\\
\textsuperscript{72} Department of Astronomy and Astrophysics, University of Chicago, 5640 South Ellis Avenue, Chicago, IL 60637, USA\\
\textsuperscript{73} Department of Physics and Astronomy, University of Missouri, 5110 Rockhill Road, Kansas City, MO 64110, USA\\
\textsuperscript{74} Enrico Fermi Institute, University of Chicago, 5640 South Ellis Avenue, Chicago, IL 60637, USA\\
\textsuperscript{75} Institute of Cosmology \& Gravitation, University of Portsmouth, Dennis Sciama Building, Portsmouth, PO1 3FX, UK\\
\textsuperscript{76} Department of Physics, University of Chicago, 5640 South Ellis Avenue, Chicago, IL 60637, USA\\
\textsuperscript{77} Department of Physics and McGill Space Institute, McGill University, 3600 Rue University, Montreal, Quebec H3A 2T8, Canada\\
\textsuperscript{78} School of Mathematics, Statistics \& Computer Science, University of KwaZulu-Natal, Durban, South Africa\\
\textsuperscript{79} University of Chicago, 5640 South Ellis Avenue, Chicago, IL 60637, USA\\
\textsuperscript{80} Jet Propulsion Laboratory, California Institute of Technology, Pasadena, CA 91011, USA\\
\textsuperscript{81} Astronomy Unit, Department of Physics, University of Trieste, via Tiepolo 11, I-34131 Trieste, Italy\\
\textsuperscript{82} INAF-Osservatorio Astronomico di Trieste, via G. B. Tiepolo 11, I-34143 Trieste, Italy\\
\textsuperscript{83} Institute for Fundamental Physics of the Universe, Via Beirut 2, 34014 Trieste, Italy\\
\textsuperscript{84} Cornell University, Ithaca, NY 14853, USA\\
\textsuperscript{85} Hamburger Sternwarte, Universit\"at Hamburg, Gojenbergsweg 112, 21029 Hamburg, Germany\\
\textsuperscript{86} School of Mathematics and Physics, University of Queensland, Brisbane, QLD 4072, Australia\\
\textsuperscript{87} Institute of Particle and Nuclear Studies (IPNS), High Energy Accelerator Research Organization (KEK), Tsukuba, Ibaraki 305-0801, Japan\\
\textsuperscript{88} International Center for Quantum-field Measurement Systems for Studies of the Universe and Particles (QUP), High Energy Accelerator Research Organization (KEK), Tsukuba, Ibaraki 305-0801, Japan\\
\textsuperscript{89} Canadian Institute for Advanced Research, CIFAR Program in Gravity and the Extreme Universe, Toronto, ON, M5G 1Z8, Canada\\
\textsuperscript{90} Department of Astrophysical and Planetary Sciences, University of Colorado, Boulder, CO 80309, USA\\
\textsuperscript{91} Departments of Statistics and Data Sciences, University of Texas at Austin, Austin, TX 78757, USA\\
\textsuperscript{92} NSF-Simons AI Institute for Cosmic Origins, University of Texas at Austin, Austin, TX 78757, USA\\
\textsuperscript{93} Institute of Cosmology and Gravitation, University of Portsmouth, Dennis Sciama Building, Burnaby Road, Portsmouth, PO1 3FX, UK\\
\textsuperscript{94} Harvey Mudd College, 301 Platt Boulevard, Claremont, CA 91711, USA\\
\textsuperscript{95} Institute of Cosmology and Gravitation, University of Portsmouth, Portsmouth, PO1 3FX, UK\\
\textsuperscript{96} Institute of Space Sciences (ICE, CSIC),  Campus UAB, Carrer de Can Magrans, s/n,  08193 Barcelona, Spain\\
\textsuperscript{97} European Southern Observatory, Karl-Schwarzschild-Str., DE-85748 Garching b. M\"unchen, Germany\\
\textsuperscript{98} CSIRO Space \& Astronomy, PO Box 1130, Bentley WA 6102, Australia\\
\textsuperscript{99} D\'epartement de Physique, Universit\'e de Montr\'eal, Succ. Centre-Ville, Montr\'eal, Qu\'ebec, H3C 3J7, Canada\\
\textsuperscript{100} Department of Physics, University of Illinois Urbana-Champaign, 1110 W. Green Street, Urbana, IL 61801, USA\\
\textsuperscript{101} Santa Cruz Institute for Particle Physics, Santa Cruz, CA 95064, USA\\
\textsuperscript{102} Department of Physics, University of California, Berkeley, CA 94720, USA\\
\textsuperscript{103} Center for Astrophysics \textbar\ Harvard \& Smithsonian, 60 Garden Street, Cambridge, MA 02138, USA\\
\textsuperscript{104} Department of Physics, University of California, One Shields Avenue, Davis, CA 95616, USA\\
\textsuperscript{105} Australian Astronomical Optics, Macquarie University, North Ryde, NSW 2113, Australia\\
\textsuperscript{106} Lowell Observatory, 1400 Mars Hill Rd, Flagstaff, AZ 86001, USA\\
\textsuperscript{107} Physics Division, Lawrence Berkeley National Laboratory, Berkeley, CA 94720, USA\\
\textsuperscript{108} Centre for Gravitational Astrophysics, College of Science, The Australian National University, ACT 2601, Australia\\
\textsuperscript{109} The Research School of Astronomy and Astrophysics, Australian National University, ACT 2601, Australia\\
\textsuperscript{110} Departamento de F\'isica Matem\'atica, Instituto de F\'isica, Universidade de S\~ao Paulo, CP 66318, S\~ao Paulo, SP, 05314-970, Brazil\\
\textsuperscript{111} Centre for Extragalactic Astronomy, Durham University, South Road, Durham DH1 3LE, UK\\
\textsuperscript{112} Institute for Computational Cosmology, Durham University, South Road, Durham DH1 3LE, UK\\
\textsuperscript{113} George P. and Cynthia Woods Mitchell Institute for Fundamental Physics and Astronomy, and Department of Physics and Astronomy, Texas A\&M University, College Station, TX 77843, USA\\
\textsuperscript{114} Kavli Institute for Astrophysics and Space Research, Massachusetts Institute of Technology, 77 Massachusetts Avenue, Cambridge, MA~02139, USA\\
\textsuperscript{115} LPSC Grenoble - 53, Avenue des Martyrs 38026 Grenoble, France\\
\textsuperscript{116} Instituci\'o Catalana de Recerca i Estudis Avan\c cats, E-08010 Barcelona, Spain\\
\textsuperscript{117} Dunlap Institute for Astronomy \& Astrophysics, University of Toronto, 50 St. George Street, Toronto, ON, M5S 3H4, Canada\\
\textsuperscript{118} David A. Dunlap Department of Astronomy \& Astrophysics, University of Toronto, 50 St. George Street, Toronto, ON, M5S 3H4, Canada\\
\textsuperscript{119} Materials Sciences Division, Argonne National Laboratory, 9700 South Cass Avenue, Lemont, IL 60439, USA\\
\textsuperscript{120} Observat\'orio Nacional, Rua Gal. Jos\'e Cristino 77, Rio de Janeiro, RJ - 20921-400, Brazil\\
\textsuperscript{121} School of Physics and Astronomy, University of Minnesota, 116 Church Street SE Minneapolis, MN 55455, USA\\
\textsuperscript{122} Department of Physics, Case Western Reserve University, Cleveland, OH 44106, USA\\
\textsuperscript{123} Brookhaven National Laboratory, Upton, NY 11973, USA\\
\textsuperscript{124} Universit\'e Paris-Saclay, CNRS, Institut d'Astrophysique Spatiale, 91405, Orsay, France\\
\textsuperscript{125} INAF - Osservatorio Astronomico di Trieste, via G. B. Tiepolo 11, 34143 Trieste, Italy\\
\textsuperscript{126} IFPU - Institute for Fundamental Physics of the Universe, Via Beirut 2, 34014 Trieste, Italy\\
\textsuperscript{127} Instituto de F\'\i sica, UFRGS, Caixa Postal 15051, Porto Alegre, RS - 91501-970, Brazil\\
\textsuperscript{128} Astronomy Unit, Department of Physics, University of Trieste, via Tiepolo 11, 34131 Trieste, Italy\\
\textsuperscript{129} INFN - National Institute for Nuclear Physics, Via Valerio 2, I-34127 Trieste, Italy\\
\textsuperscript{130} ICSC - Italian Research Center on High Performance Computing, Big Data and Quantum Computing, Italy\\
\textsuperscript{131} Liberal Arts Department, School of the Art Institute of Chicago, 112 South Michigan Avenue, Chicago, IL 60603, USA \\
\textsuperscript{132} Department of Astronomy, University of Michigan, 1085 S. University Ave, Ann Arbor, MI 48109, USA\\
\textsuperscript{133} Three-Speed Logic, Inc., Victoria, B.C., V8S 3Z5, Canada\\
\textsuperscript{134} Physics Department, Lancaster University, Lancaster, LA1 4YB, UK\\
\textsuperscript{135} Department of Physics, Faculty of Science, Chulalongkorn University, 254 Phayathai Road, Pathumwan, Bangkok 10330, Thailand\\
\textsuperscript{136} Computer Science and Mathematics Division, Oak Ridge National Laboratory, Oak Ridge, TN 37831, USA\\
\textsuperscript{137} Space Science and Engineering Division, Southwest Research Institute, San Antonio, TX 78238, USA\\
\textsuperscript{138} Department of Physics and Astronomy, Michigan State University, East Lansing, MI 48824, USA\\
\textsuperscript{139} Department of Astronomy \& Astrophysics, University of Toronto, 50 St George St, Toronto, ON, M5S 3H4, Canada

\end{document}